\documentclass[journal=apchd5,manuscript=letter]{achemso}

\usepackage{chemformula} % Formula subscripts using \ch{}
\usepackage[T1]{fontenc} % Use modern font encodings
\usepackage{amssymb}
\usepackage{array}
\usepackage{booktabs}
\usepackage{makecell} % For folding text in table cells
\usepackage{hyperref}
\usepackage{soul}
\usepackage{float}  % Add this to your preamble if not already present
\author{Mengdi Sun}
\email{mengdis@vt.edu}
\author{Ata Shakeri}
\author{Arvin Keshvari}
\author{Dimitrios~Giannakopoulos}
\affiliation{Bradley Department of Electrical and Computer Engineering, Virginia Tech, Blacksburg, Virginia 24060, United States}
\author{Qing Wang}
\author{Wei-Ting~Chen}
\affiliation{SNOChip Inc., Princeton, New Jersey 08540, United States}
\author{Steven~G.~Johnson}
\affiliation{Department of Mathematics, Massachusetts Institute of Technology, Cambridge, Massachusetts 02138, United States}
\author{Zin~Lin}
\affiliation{Bradley Department of Electrical and Computer Engineering, Virginia Tech, Blacksburg, Virginia 24060, United States}

\title[An \textsf{achemso} demo]
  {Scalable freeform optimization of wide-aperture 3D metalenses by zoned discrete axisymmetry}


\begin{document}

\begin{tocentry}
\includegraphics[width=\linewidth]{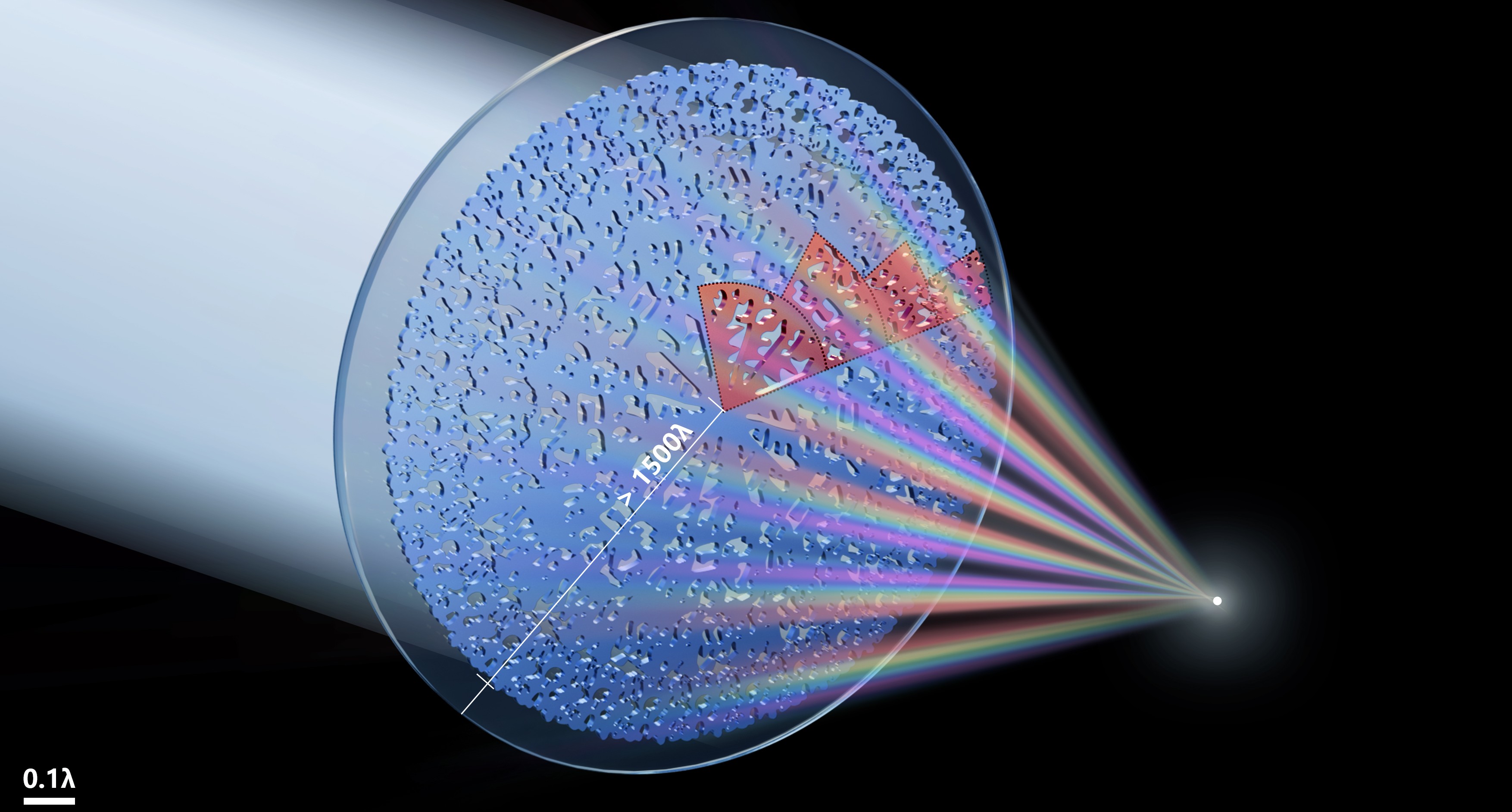}

For Table of Contents Use Only.

An artistic rendition of a wide-aperture , poly-achromatic, 3D metalens with nano-scale freeform features.

\end{tocentry}

\begin{abstract}
\label{sec:Abs}
We introduce a novel framework for design and optimization of 3D freeform metalenses that attains nearly linear scaling of computational cost with diameter, by breaking the lens into a sequence of radial ``zones'' with $n$-fold discrete axisymmetry, where $n$ increases with radius. This allows vastly more design freedom than imposing continuous axisymmetry, while avoiding the compromises of the locally periodic approximation (LPA) or scalar diffraction theory. Using a GPU-accelerated finite-difference time-domain (FDTD) solver in cylindrical coordinates, we perform full-wave simulation and topology optimization within each supra-wavelength zone. We validate our approach by designing millimeter and centimeter-scale, poly-achromatic, 3D freeform metalenses which outperform the state of the art. By demonstrating the scalability and resulting optical performance enabled by our ``zoned discrete axisymmetry'' (ZDA) and supra-wavelength domain decomposition, we highlight the potential of our framework to advance large-scale meta-optics and next-generation photonic technologies. 
\end{abstract}

\section{Introduction}
\label{sec:Intro}
While metasurfaces have been hailed as a revolutionary platform for developing ultra-compact imaging devices~\citep{Khorasaninejad:16,Chen:18,Rubin:19,Li:21,Li:22}, their performance is often hampered by the computational challenges of modeling irregular surfaces that can be many thousands of wavelengths in diameter, often necessitating severe approximations to the wave physics and/or restrictions on the designs. For example, many authors have employed the locally-periodic approximation (LPA), which breaks the surface into a set of ``meta-atom'' unit cells (typically sub-wavelength $<\lambda$) with approximate periodic boundary conditions that allow each unit cell to be simulated independently~\citep{Lin:19,Pestourie:18,Chung:23,Bayati:20}. The period-zero limit of LPA is to approximate the surface in terms of locally uniform surfaces, leading to scalar-diffraction theory and local ray-optics models~\citep{O'Shea:04,Voronovich:99}. Other authors have broken the surface into small subdomains with approximate absorbing boundary conditions, effectively neglecting long-range interactions, to facilitate parallel decoupled simulations~\citep{Zhou:24,Lin:19(OE),Lin:21(APL),Zhelyeznyakov:23}, but the overall cost still scales with the metasurface area. In general, approximate solvers such as LPA may limit the attainable performance~\citep{Chung:20}. An alternative approach has been to restrict the metasurface design space to axisymmetric structures~\citep{Lin:21(APL),Christiansen:20}, which allows the computational cost of full-wave Maxwell simulations to scale nearly linearly with the diameter of the metasurface at the price of greatly limiting the designs---for single-layer lithography, the designs are restricted to sets of concentric rings.  The most obvious relaxation of continuous axisymmetry would be to \emph{discrete} $n$-fold axisymmetry, but in this case the computational domain would be a circular wedge whose area still increases quadratically with diameter. We would like to match the nearly \emph{linear} scaling of computational cost with diameter of axisymmetric designs, while retaining full-wave physical modeling of large (many-wavelength) domains, and at the same time support exploration of a large design space of complicated (non-axisymmetric) surface patterns, ideally by freeform topology optimization (TopOpt)~\citep{Lin:18,Lin:19,Lin:19(OE)}.

Our solution is to introduce a new type of axisymmetry, which we refer to as ``zoned discrete axisymmetry'' (ZDA), in which we employ $n$-fold discrete axisymmetry where $n$ increases roughly proportionally to radius, divided into a set of discrete radial zones, giving a total design area that scales linearly with diameter (Section ``Zoned discrete axisymmetry (ZDA)'').  For example, a millimeter or centimeter-scale metalens is divided into just a few concentric zones $\{\Omega_i, i\lesssim 30\}$, each possessing a discrete ($n_i$-fold) axial symmetry and a radial width $R_i \gg \lambda$. Such a structure can be solved by a wide variety of computational methods, with fullwave simulation of each zone and approximate absorbing boundary conditions between zones that suppresses near-field inter-zone interactions, and given a differentiable solver one can apply large-scale optimization techniques such as TopOpt.
In this paper, we employ a differentiable finite-difference time-domain (FDTD) solver in cylindrical coordinates, capable of rapidly optimizing wide-aperture 3D freeform metalenses (Fig.~\ref{fig:1}) without invoking LPA. We further accelerate our solver with a GPU, achieving a throughput of one billion voxels/sec, while the domain decomposition approach also allows a straightforward distribution and parallelization of different domains over multiple GPUs. We report proof-of-concept, topology-optimized freeform large-area metalens designs (Section ``Results and Discussion''): a 0.8-cm-diameter monochromatic LWIR metalens with NA = 0.3 and $\sim$ 58$\%$ focusing efficiency, a 1-mm-diameter RGB-achromatic metalens with NA = 0.8 and an average focusing efficiency of $\sim$ 33$\%$, and a 2-mm-diameter poly-achromatic metalens optimized for 6 wavelengths with NA = 0.3 and an average focusing efficiency of $\sim$ 12$\%$.  We believe that the ZDA approach will enable many exciting future applications of metasurfaces that go beyond the limitations of LPA or axisymmetric structures (Section ``Summary and Outlook''), especially for problems where the optimal design is far from the conventional Fresnel-like lenses where LPA worked well \citep{Engelberg:20,Arbabi:23,Lalanne:17}.

\section{Zoned discrete axisymmetry (ZDA)}
\label{sec:Features}
Metalenses are critical components of any metaoptical imaging systems. While it may be desirable to simulate a wide-aperture metalens in its entirety without any approximation~\citep{Xue:23,Minkov:24}, we emphasize that it is not necessary to do so. First of all, it is only natural to impose some degree of axial (rotational) symmetry on the metalens geometry, which focuses light to a point on the optical axis. Empirical evidence \citep{Lin:19} indicates that even in the absence of any \textit{explicitly} imposed symmetries, axisymmetric patterns will automatically emerge out of the freeform inverse design of an entire metalens. Therefore, it is prudent to proactively minimize unnecessary computational burdens by implementing symmetry reductions before performing any simulation or optimization. Secondly, in most free-space flat-optics devices, where light comes from the free space on one side and transmits to the free space on the other side over the same area, light propagation and throughput is primarily ``local'', while possible lateral spread or long-range (non-local) transverse interactions inside the device are fundamentally limited by the device thickness \citep{Miller:23,Li:22(light)}. Therefore, it is physically justifiable to resort to some form of domain decomposition, as long as the transverse dimension of each domain is considerably larger than its thickness. Indeed, we have shown that decomposing a large-area metalens into a few supra-$\lambda$ ($\gg \lambda$) zones, each terminated by PMLs, serve to reduce stray light errors in the far field to practically negligible proportions (less than 1\% error, compared to a full simulation, when utilizing >~10$\lambda$-wide zones)~\citep{Lin:19(OE)}. 

\begin{figure}[htbp]
    \centering
    \includegraphics[width=\linewidth]{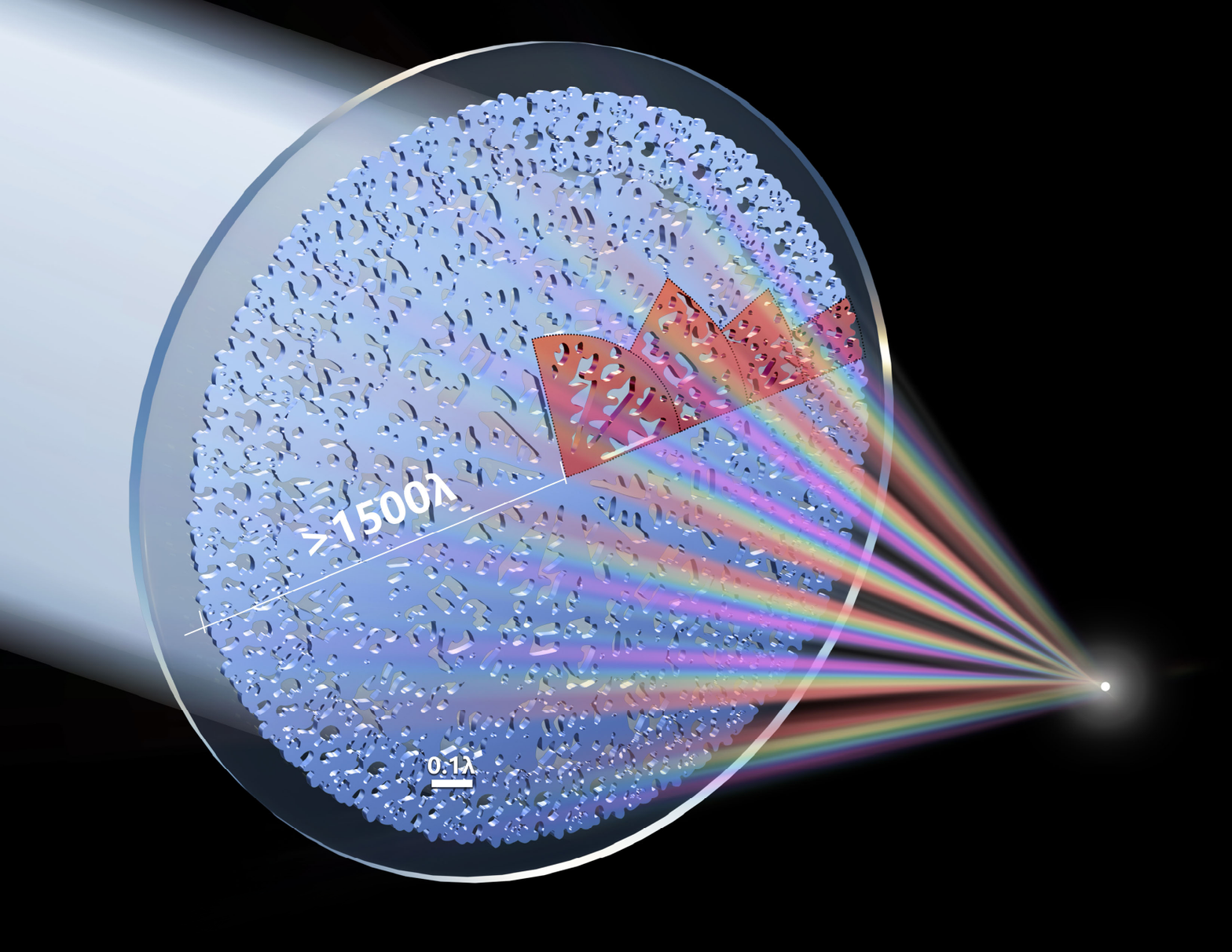}
    \caption{An artistic rendition (not drawn to scale) of a wide-aperture (diameter > 3000$\lambda$), poly-achromatic, 3D metalens with nano-scale (<0.1$\lambda$) freeform features. Only the shaded areas need to be simulated and optimized, following a zoned discrete axisymmetry, where concentric supra-$\lambda$ radial zones have increasing azimuthal symmetry (see Fig.~\ref{fig:2}).}
    \label{fig:1}
\end{figure}

These considerations motivate us to simulate Maxwell's equations in cylindrical coordinates $(r,\phi,z)$ instead of the familiar Cartesian coordinates, and to decompose a large-area metalens design into radially-concentric, supra-$\lambda$ zones, where the computation in each domain shall be further reduced by azimuthally-periodic ($\phi$-periodic) Bloch boundary conditions under an $n$-fold symmetry. The degree of axial symmetry, $n \in \mathbb{Z}^+$, will be set to increase approximately linearly with radius from the inner zones to the outer zones, which serves to \textit{fractionalize} the azimuthal period at larger radii (see Fig.~\ref{fig:1}) and therefore constrains the simulation area to scale linearly with the radius (instead of a much more expensive quadratic scaling). We note that while our prior works have reported on continuously axisymmetric designs~\citep{Christiansen:20} (i.e., $n\rightarrow\infty$), discrete axial symmetry enables a dramatically enlarged design space, introducing fully two- or three-dimensional degrees of freedom (DOFs) in contrast to the limited one- or two- dimensional DOFs available to continuous axisymmetry. In Section ``Centimeter-scale metalens at long-wave infrared'', we will show that discrete axisymmetry is critical to significantly increasing the metalens performance beyond what is achievable with continuous axisymmetry, whose performance struggles under binarization. We would like to note that a previous work~\citep{Chung:23} introduces \textit{ad hoc} periodic azimuthal variations (i.e., without any simulation or optimization) as a (crude) mean-field effective-index approximation of a greyscale, continuously axisymmetric metalens design. In contrast, our work enables rigorous simulation and optimization of freeform discrete axisymmetric azimuthal variations using fullwave Maxwell equations. 

\subsection{Differentiable FDTD in cylindrical coordinates}
\label{sec:Diff_FDTD}
\begin{figure}[htbp]
    \centering
    \includegraphics[width=\linewidth]{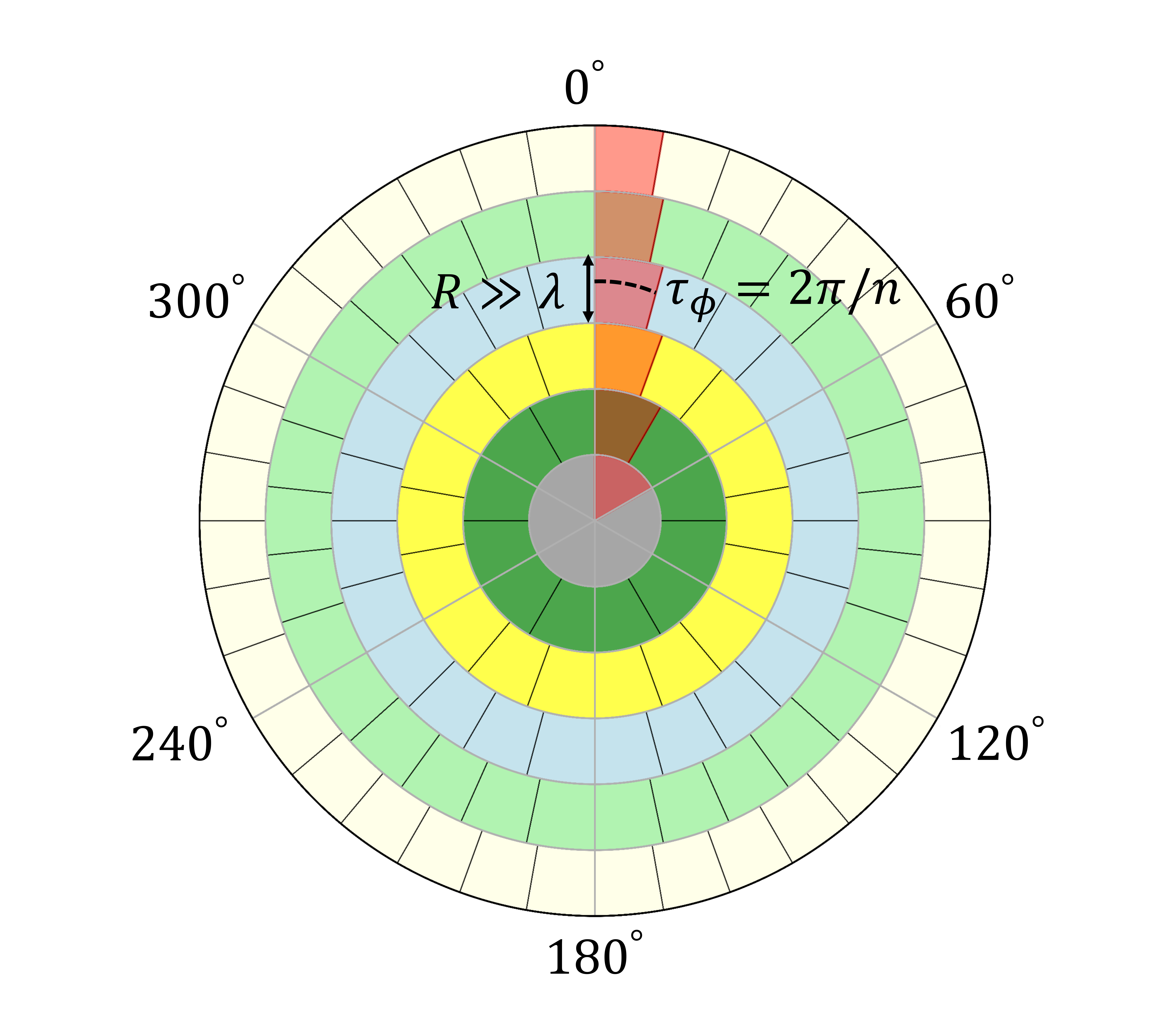}
    \caption{Our zoned discrete axisymmetry (ZDA) scheme. Only the red-shaded domains are simulated and optimized. For each zone, the radial width $R$ is much larger than $\lambda$ and the azimuthal period $\tau_{\phi}$ equals to 2$\pi$ divided by the degree of axial symmetry $n$.}
    \label{fig:2}
\end{figure}

For an $n$-fold symmetric permittivity distribution $\varepsilon(r,\phi,z)=\varepsilon(r,\phi+\tau_\phi,z)$ with periodicity $\tau_\phi = 2 \pi / n$ (Fig.~\ref{fig:2}), Bloch theorem allows us to re-write the electric and magnetic fields:
\begin{align}
    \mathbf{E} &= \sum_m \mathbf{u}_m(r,\phi,z) e^{im\phi}  \label{eq:1} \\
    \mathbf{H} &= \sum_m \mathbf{v}_m(r,\phi,z) e^{im\phi}   \label{eq:2}
\end{align}
where $\mathbf{u}_m(r,\phi,z)=\mathbf{u}_m(r,\phi+\tau_\phi,z),~ \mathbf{v}_m(r,\phi,z)=\mathbf{v}_m(r,\phi+\tau_\phi,z)$, and $m$ represents the azimuthal mode number (or Bloch angular momentum). 

The reduced Maxwell equations for $\mathbf{u}_m$ and $\mathbf{v}_m$ are then given by:
\begin{align}
    \mathcal{D}_m~\mathbf{u}_m &= -\mu \frac{\partial \mathbf{v}_m}{\partial t} \label{eq:3} \\
    \mathcal{D}_m~\mathbf{v}_m &= \varepsilon \frac{\partial \mathbf{u}_m}{\partial t} + \mathbf{J}_m \label{eq:4}
\end{align}
where
\begin{align}
{\cal D}_m &= 
\begin{pmatrix}
0 		& 	- {\partial \over \partial z} 		& 	{im \over r} + {1 \over r}{\partial \over \partial \phi} \\
 \partial \over \partial z	&	0			&	- {\partial \over \partial r} \\
-{i m \over r} - {1 \over r}{\partial \over \partial \phi}	&	{1 \over r}{\partial \over \partial r} r	&	0 
\end{pmatrix} \label{eq:5}
\end{align}
We discretize these equations with a staggered Yee grid in space and a leap-frog updating scheme in time~\citep{Taflove:05}. Special care is needed to discretize the center $r=0$, which represents a numerical (but not physical) singularity. We remove this singularity by applying Faraday and Ampere loop-integrals around the center voxels~\citep{Taflove:05}. We surround each zone with stretched-coordinate PMLs~\citep{Taflove:05} in the $(r,z)$ directions. For the zone that contains $r=0$, we only include an outer radial PML layer, while we include both outer and inner PMLs for all other zones. We note that the radial $r$-PMLs are qualitatively different from the vertical $z$-PMLs because of the \textit{explicit} appearance of the stretched coordinate $r'= r + i {\zeta_r(r) \over \omega}$ in the Maxwell equations themselves, where $\zeta_r(r)$ is a PML stretch factor~\citep{Taflove:05} and $\omega$ is the angular frequency. Consequently, care must be taken to regularize the $r$-PMLs to avoid numerical divergence~\citep{Teixeira:99}. Finally, we translate our algorithm into a GPU program using Julia CUDA v5.4 module. The algorithmic structure of FDTD is fairly susceptible to single-program multiple-data (SPMD) parallelism on a GPU; our preliminary GPU implementation, without any in-depth occupancy maximization, yields a FDTD throughput of $\sim$1 billion voxels per second on an Nvidia A100 GPU, while we expect >30 billion voxels/sec after thorough GPU optimizations~\citep{Minkov:24} (and even higher speedups on newer GPU models). 

Freeform inverse design requires the ability to compute the gradient of an objective function $f$ with respect to the full $\varepsilon(r,\phi,z)$ profile. We render our software fully differentiable by implementing a hybrid time/frequency-domain adjoint sensitivity analysis~\citep{Hammond:22}  within the Julia automatic differentiation (AD) environment (based on Julia Zygote AD~\citep{Zygote.jl-2018}). The adjoint-AD integration approach is especially important to allow  efficient \textit{backpropagation} from an arbitrary objective $f$, which may be a complicated function of the far field, involving, for example, image-processing algorithms~\citep{Lin:21(Nanophot),Lin:22(OE)}. We note that the multi-frequency forward and adjoint fields can be efficiently extracted \textit{all at once} from a single forward or adjoint time-domain simulation by discrete-time Fourier transforms (DTFT). To accomplish this requires excitations using broadband adjoint sources, which are spatio-temporally inseparable superpositions of suitably-chosen temporal pulses modulated by spatially-varying adjoint forces (e.g., ${\partial f \over \partial \mathbf{E}})$~\citep{Hammond:22}. We will report a detailed implementation of our adjoint-differentiable discrete axisymmetric FDTD software elsewhere.

\section{Results and Discussion}
\label{sec:R&D}
The 3D freeform metalens inverse design is achieved by adjoint gradient based topology optimization. The permittivity $\varepsilon$ is parametrized by:
\begin{align}
    \varepsilon(\vec{p}) = \vec{p} ~ (\varepsilon_{h} - \varepsilon_{0}) + \varepsilon_{0} \label{eq:6}
\end{align}
where $\vec{p} \in [0,1]^N$ is the DOF array representing the discretized grid, with $N$ grid points, over the design region, $\varepsilon_{h}$ is the permittivity of the high-index metalens material, and $\varepsilon_{0}$ is the permittivity of the background material (typically vacuum $\varepsilon_{0}$ = 1). As an example, we will primarily consider maximizing the far field intensity $I(\vec{p})$ at the designated focal spot of a metalens under a normally incident plane wave. (Note that a normally incident plane wave can be decomposed into $m=\pm1$ components in the cylindrical coordinates~\citep{Lin:21(APL),Christiansen:20}.) To realize a poly-achromatic or achromatic lens, we will maximize the minimum of multiple focal intensities $\min_i ~\{I(\vec{p},\omega_i), i=1,2,3,...\}$, which can be formulated into an epigraph form~\citep{Boyd:04}:
\begin{align}
    \max_{\vec{p},q} &\quad q \label{eq:7} \\
    q \leq~ &I(\vec{p},\omega_i), \quad i = 1,2,3, ... \label{eq:8}
\end{align}
To create experimentally feasible designs, we apply the standard binarization and conic filters to $\vec{p}$ during topology optimization~\citep{jensen2011topology}.

\subsection{Centimeter-scale metalens at long-wave infrared}
\label{sec:mono}
\begin{figure}[htbp]
    \centering
    \includegraphics[width=\linewidth]{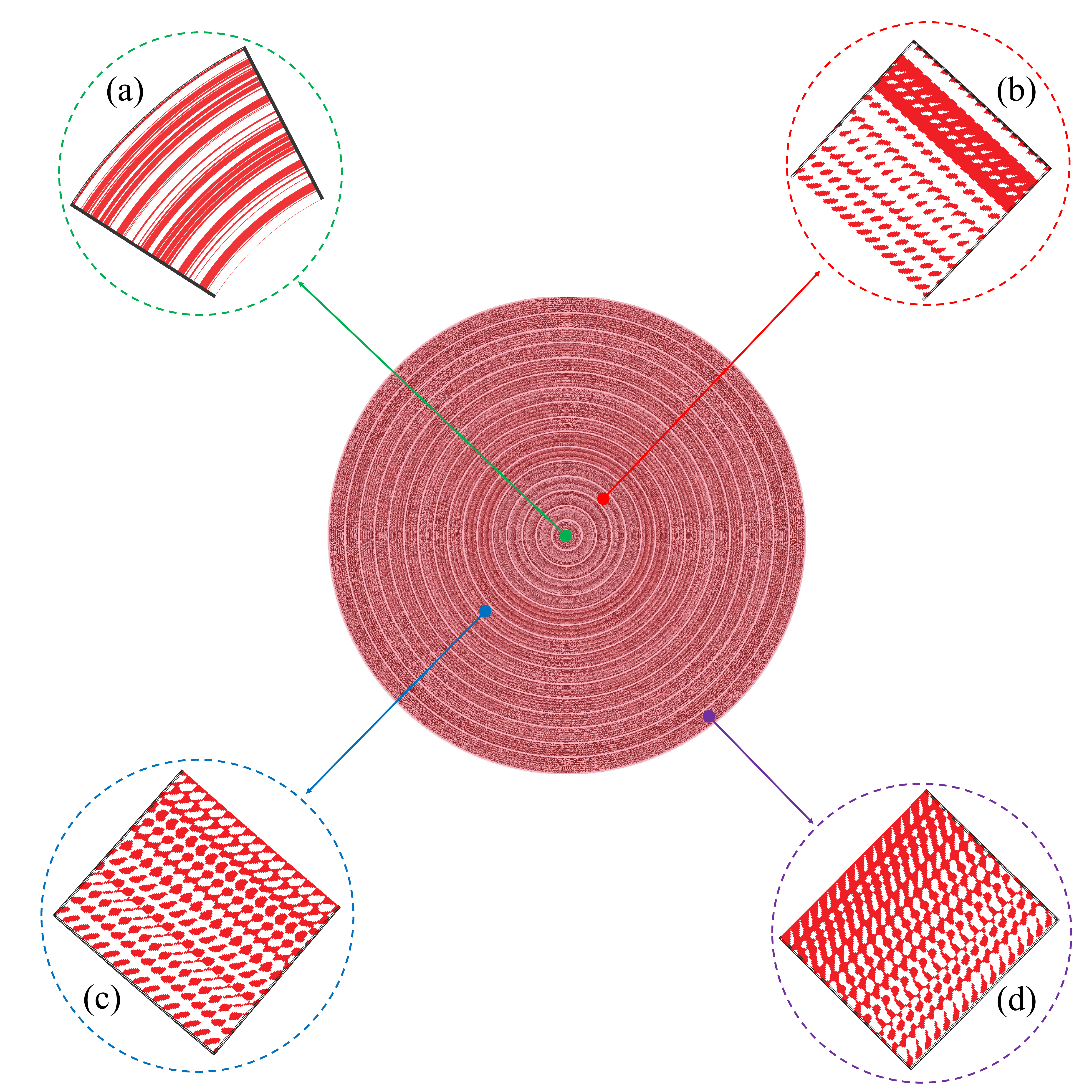}
    \caption{The schematic of the structure of the inverse-designed 8-mm diameter monochromatic LWIR metalens. Fig.~\ref{fig:3}(a) - (d) are the zoomed-in insets of the 1st (0-25$\lambda$), the 4th (75$\lambda$-100$\lambda$), the 8th (175$\lambda$-200$\lambda$) and the 16th (375$\lambda$-400$\lambda$) zones respectively.}
    \label{fig:3}
\end{figure}

To demonstrate the ability of discrete axisymmetric inverse design, we first optimize a monochromatic metalens with 800$\lambda$ or 0.8 cm in diameter (operating at a long-wave infrared $\lambda=10\mathrm{\mu m}$) and compare it against an optimized continuous axisymmetric design. The numerical aperture (NA) is chosen to be 0.3 (which corresponds to a focal length of 1.27cm).  We use silicon ($n_{Si}$ = 3.42 at $\lambda = 10\mathrm{\mu m}$) for both the metalens layer (thickness = 6$\mathrm{\mu m}$ or 0.6$\lambda$) and the substrate. The design region is radially decomposed into 16 supra-$\lambda$ zones, each with a radial width of 25$\lambda$. To facilitate azimuthally invariant focusing, stray diffractions into higher-order angular momenta ($m'=m + k n, k = \pm1, \pm2, ...$) should be avoided, hence we choose a sufficiently high $n$ (the degree of axial symmetry) for each zone (such that $r\tau_\phi < \lambda$) to ensure that higher diffraction orders cannot propagate. Specifically, we choose the first zone to be continuously axisymmetric ($n\rightarrow\infty$) while the azimuthal period for the $\ell$-th zone ($\ell = 2, ..., 16$) is set to be $\tau^{(\ell)}_\phi = 2\pi / 720\ell$. To ensure simulation convergence, a minimum spatial resolution of $\lambda \over 15 n_{index}$, where $n_{index}$ is the highest refractive index of all materials in the system, is required~\citep{Shin:12,Kunz:93} along with the Courant-Friedrichs-Lewy condition~\citep{Courant:67}. Here we set the discretization steps $\delta r = \delta z = \lambda/50$ and $r \delta \phi < 0.021 \lambda$. To discourage the appearance of fine features and to make our design manufacturable, we applied a conic filter with a radius of $0.08\lambda$ during topology optimization. The optimized design is shown in Fig.~\ref{fig:3}, where a total of 1.26 billion pixels were configured to create a fully-freeform metalens (red pixels indicate Si and white indicates vacuum).

\begin{figure}[htbp]
    \centering
    \includegraphics[width=\linewidth]{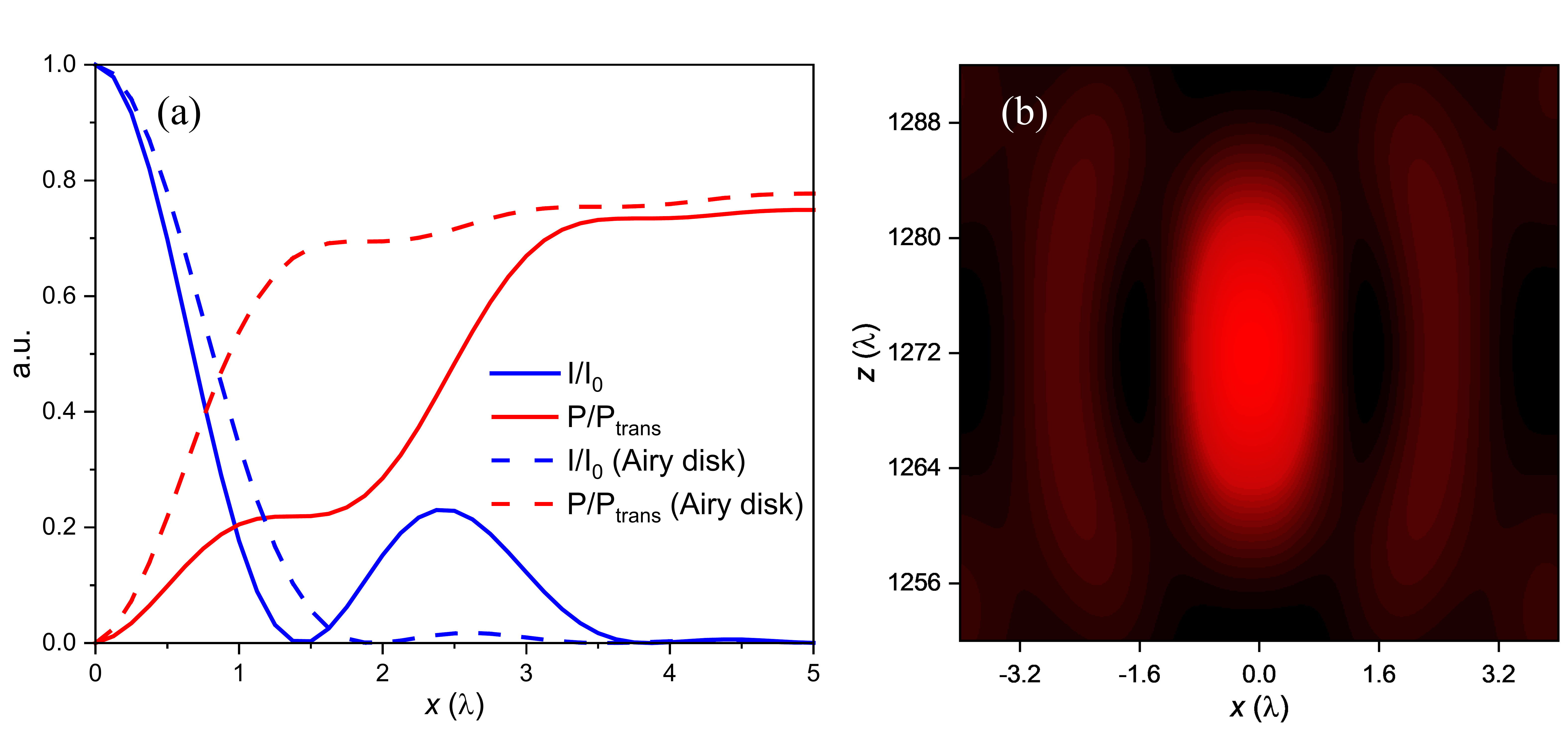}
    \caption{Focal spot properties of the 8-mm-diameter and NA = 0.3 metalens at a wavelength of 10 $\mu$m. (a) The simulated point spread function (PSF) and (b) the focal intensity distribution in the $xz$-plane of the monochromatic LWIR metalens.}
    \label{fig:4}
\end{figure}

The point spread function (PSF) of the optimized metalens is shown in Fig.~\ref{fig:4}(a) (with the ideal Airy disk as a reference). The full width at half maximum (FWHM) of the PSF is 1.34$\lambda$, even outperforming the ideal Airy disk, whose FWHM is 1.6$\lambda$, indicating that our design is a super-oscillatory lens~\citep{Chen:19}. It is found that our design shows a transmittance of $\approx80\%$, which is surprisingly high considering that a uniform Si-air interface has a reflectance of $\approx30\%$. The high transmittance suggests that the optimization created nanostructures whose effective medium index lies between Si and air, significantly mitigating the reflections. The absolute focusing efficiency (defined as the fraction of the total incident power within 3 FWHMs from the center of the focal spot~\citep{Arbabi:15,Engelberg:22}) is found to be 58.4$\%$, whereas the relative focusing efficiency (normalized against the transmitted power) is 73.5$\%$. To validate the concept of discrete axial symmetry, we also topology-optimized an alternative metalens design with a purely continuous axial symmetry but the same diameter, NA, and material parameters. We found that the continuous axisymmetric metalens struggled to achieve a high transmittance during binarization, ultimately resulting in a focusing efficiency of 25.1$\%$ (which is 57\% smaller than that of the discrete axisymmetric design). Therefore, the significantly superior performance of the discrete axisymmetric design can be attributed to the introduction of additional DOFs along the azimuthal direction (roughly, in our optimization setting, there are 10 times more DOFs under the discrete symmetry than under the continuous symmetry), which allowed the optimization to discover effective-medium nanostructures. Notably, one optimization iteration (2 FDTD runs) of the 0.8-cm metalens design finishes in 5 minutes on 8 Nvidia A100 GPUs, and the entire inverse design took $\sim$ 159 hours.

To further characterize our metalens design, we also \textit{rigorously} compute the Strehl ratio (SR), which is defined as the peak-to-peak ratio between the metalens focal-spot intensity ($I$) and an ideal Airy disk ($I_\text{Airy}$):
\begin{align}
    \mathrm{SR} &= {I\over I_\text{Airy}} = {I \lambda^2 F^2 \over P_\text{trans} A},
    \label{eq:9}
\end{align}
where $F$ is the focal length, and the Airy disk should be carefully normalized with respect to the total transmitted power flux $P_\text{trans}$ integrated just above the \textit{entire} metalens area $A$. We report an SR of 0.5, which is relatively low because we did not directly maximize the SR in our inverse design but the un-normalized focal intensity instead. As a consequence, the optimization created a so-called super-oscillatory lens~\citep{Chen:19,Shim:20}, with a sub-Rayleigh focal spot even smaller than the ideal Airy disk (Fig.~\ref{fig:4}(a)), albeit at the cost of a larger side lobe. On the other hand, we note that it is generally quite difficult to find a fair, reliable benchmark for Strehl ratio in the metalens literature, because of a widespread inconsistency~\citep{Engelberg:22,Menon:23,Engelberg:20} in computation/characterization of SRs for existing metalens designs (e.g., many papers compute or measure $P_\text{trans}$ over a limited area in the focal plane, which overestimates the SR). Besides, the traditional gold standard of SR>0.8 loses much of its importance when the metalens is combined with a computational image-reconstruction backend~\citep{Huang:24,Carmes:24}, which can effectively correct residual aberrations under reasonable image priors~\citep{peng2016diffractive}. Most importantly, in such applications, the transmission and focusing efficiencies are better indicators of good performance, since these efficiencies represent a high number of signal photons relative to the detection noise in the raw image, which encourages a high-fidelity, noise-resilient reconstruction outcome. We observe that the low SR of our design can be attributed primarily to the larger side lobe, and much less to stray diffractions which fall far from the focal spot. (In contrast, stray diffractions are a common issue in phase-mask model diffractive elements or metasurfaces designed under sub-$\lambda$ locally-periodic unit-cell approximations.) Therefore, despite the low SR, our design would contribute almost the same number of photons to the raw image as the ideal Airy profile (when integrated over an area including the secondary side lobe). Fig.~\ref{fig:4}(a) shows the encircled power fraction as a function of the integration area around the center. Importantly, under a high photon flux (relative to the detection noise), the effects of the side lobe can be reliably reversed by computational reconstruction, which could even exploit the sub-Rayleigh focal spot for super-resolution imaging~\citep{li2014computational,gbur2019using}. 

\begin{table}[H]  % Use [H] to force placement (requires the float package)
\centering
\caption{Parameters and performance metrics of some of the state-of-the-art metalenses.}
\label{tab:1}
\resizebox{\textwidth}{!}{%
\small
\begin{tabular}{@{}c c c c c c c@{}}
\toprule
\textbf{Type} & \textbf{Material} & \textbf{Wavelength} & \textbf{NA} & \textbf{R} & \textbf{FE} & \textbf{Ref} \\ 
\midrule
3D & CSAR-62 (n=1.59) on ITO on SiO\textsubscript{2} & 405 nm & 0.6 & 67.5 $\lambda$ & 80\% & \citep{Chung:23} \\
2D & Si\textsubscript{3}N\textsubscript{4} on SiO\textsubscript{2} & 625 nm & 0.24 & 40 $\lambda$ & 20\% & \citep{Bayati:20} \\
3D & Si on SiO\textsubscript{2} & 900 nm & 0.6 & 40.5 $\lambda$ & 66\% & \citep{Zhou:24} \\
3D & Si on Si & 10.6 $\mu$m & 0.59 & 2359 $\lambda$ & 43.4\% & \citep{Hou:24} \\
3D & TiO\textsubscript{2} on SiO\textsubscript{2} & 633 nm & 0.6 & 15.8 $\lambda$ & 62\% & \citep{Sang:22} \\
3D & TiO\textsubscript{2} on SiO\textsubscript{2} & 580 nm & 0.94 & 862 $\lambda$ & 79\% & \citep{Byrnes:16} \\
3D & Si on SiO\textsubscript{2} & 1550 nm & 0.37 & 129 $\lambda$ & 82\% & \citep{Arbabi:15} \\
3D & TiO\textsubscript{2} on SiO\textsubscript{2} & 532 nm & 0.9 & 470 $\lambda$ & 42\% & \citep{Chen:17} \\
2D & nH=2 on nL=1 & $\lambda$ & 0.9 & 25 $\lambda$ & 88\% & \citep{Li:24} \\
3D & Si on SiO\textsubscript{2} & 1.3 - 1.65 $\mu$m & 0.24 & 30.3 $\lambda$ & 50\% & \citep{Shrestha:18} \\
2D & nH=2 on nL=1 & 500 nm & 0.9 & 10000 $\lambda$ & 56\% & \citep{Xue:23} \\
3D & TiO\textsubscript{2} on SiO\textsubscript{2} & 488/532/658 nm & 0.7 & 1520 $\lambda$ & 15\% & \citep{Li:22} \\
3D & TiO\textsubscript{2} on SiO\textsubscript{2} & \makecell{490/520/540/\\570/610/650 nm} & 0.3 & 1538 $\lambda$ & 8\% & \citep{Li:22} \\
3D & TiO\textsubscript{2} on SiO\textsubscript{2} & 470/548/647 nm & 0.3 & 1543 $\lambda$ & 14\% & \citep{Li:21} \\
3D & TiO\textsubscript{2} on SiO\textsubscript{2} & 488/532/658 nm & 0.7 & 1520 $\lambda$ & 11\% & \citep{Li:21} \\
3D & Si on Si & 10 $\mu$m & 0.3 & 400 $\lambda$ & 58\% & \textbf{This work} \\
3D & TiO\textsubscript{2} on SiO\textsubscript{2} & 488/532/658 nm & 0.8 & 800 $\lambda$ & 33.1\%& \textbf{This work} \\
3D & TiO\textsubscript{2} on SiO\textsubscript{2} & \makecell{490/520/540/\\570/610/650 nm} & 0.3 & 1504 $\lambda$ & 12\% & \textbf{This work} \\ 
\bottomrule
\end{tabular}%
}
\end{table}

\subsection{Millimeter-scale poly-achromatic metalenses in the visible}
\label{sec:poly}
\begin{figure}[htbp]
    \centering
    \includegraphics[width=\linewidth]{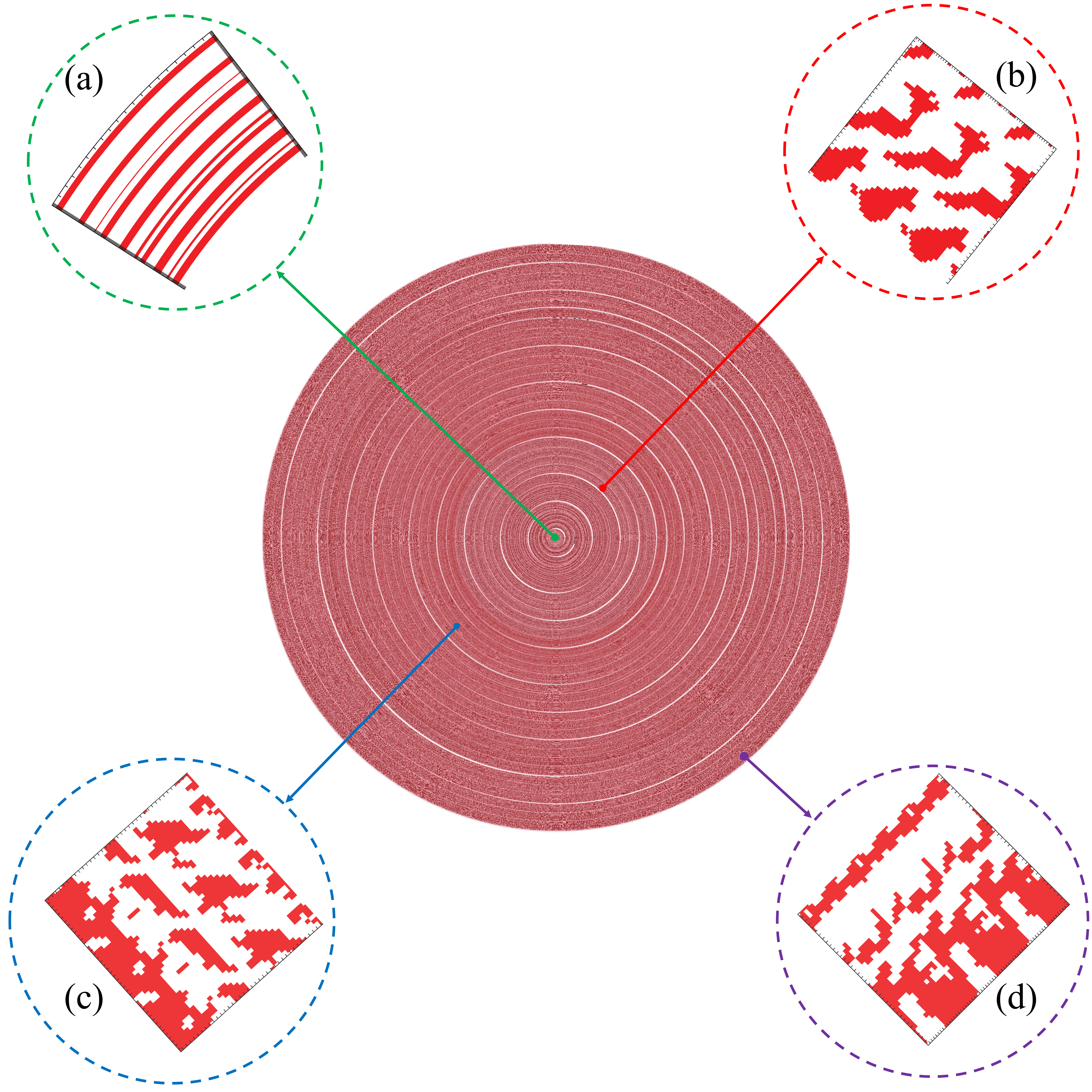}
    \caption{The schematic of the structure of the inverse-designed 1-mm diameter RGB-achromatic metalens. Fig.~\ref{fig:5}(a) - (d) are the zoomed-in insets of the 1th (0-25$\lambda$), the 8th (175$\lambda$-200$\lambda$), the 16th (375$\lambda$-400$\lambda$) and the 32th (775$\lambda$-800$\lambda$) zones respectively.}
    \label{fig:5}
\end{figure}

\begin{figure}[htbp]
    \centering
    \includegraphics[width=\linewidth]{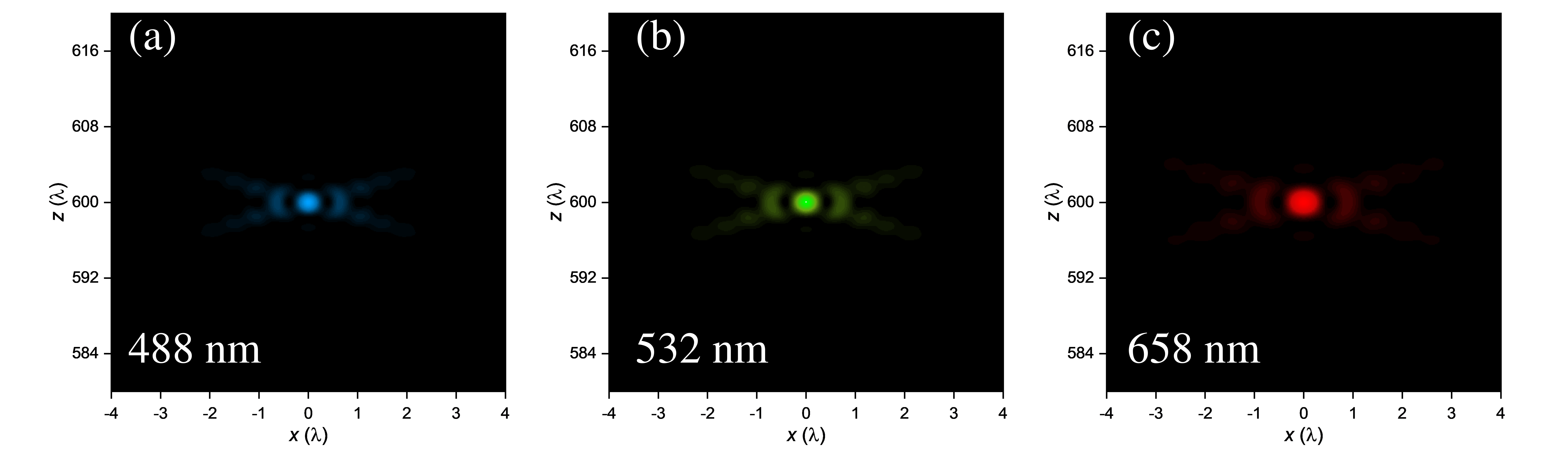}
    \caption{Simulated focal intensity distribution of the 1-mm-diameter and NA=0.8 RGB-achromatic metalens in the $xz$-plane at (a) 488 nm, (b) 532 nm and (c) 658 nm.}
    \label{fig:6}
\end{figure}

To demonstrate the versatility of our software, we optimized wide-aperture 3D freefom metalenses with poly-achromaticity; i.e., aberration corrections at multiple wavelengths---arguably, one of the most persistent and challenging problems in metalens design. First, we inverse-designed a 1.05mm diameter (1600$\lambda$) RGB-achromatic (488nm, 532nm and 658nm, corresponding to 0.74$\lambda$, 0.81$\lambda$ and $\lambda$) metalens with a high NA = 0.8 (corresponding to a focal length of 0.4mm or 600$\lambda$). Here we use 658nm-thick titanium oxide ($n_{\text{TiO}_2}$ = 2.4 at $\lambda = 658\mathrm{nm}$) on fused silica ($n_{\text{SiO}_2}$ = 1.46), which is a common material platform for visible metalenses \citep{Li:22,Li:21}. 
The metalens is radially decomposed into 32 supra-$\lambda$ zones, each with a radial width of 25$\lambda$. Continuous axial symmetry is applied in the first zone while the azimuthal period for for the $\ell$-th zone ($\ell = 2, ..., 32$) is set to be $\tau^{(\ell)}_\phi = 2\pi / 360\ell$. The spatial discretization step is set to be ${\lambda \over 36} = {\lambda \over 15 n_{index}}$. A conic filter with a radius of 0.056$\lambda$ is also applied during optimization. The optimized design is shown in Fig.~\ref{fig:5}. The far-field intensity distributions in the $xz$-plane at the three target wavelengths are shown in Fig.~\ref{fig:6}(a)-(c), which shows a perfect alignment of the focal spots (achromaticity). Due to the high numerical aperture, a much shorter depth of focus ($\sim$ 2$\lambda$) is observed compared to Fig.~\ref{fig:4}. The optimized metalens exhibits absolute focusing efficiencies of 31.6\%, 34.1\% and 33.6\% at 488nm, 532nm and 658nm respectively, while the average focusing efficiency is found to be 33.1\% overall. By comparison, the state-of-the-art high-NA (0.7), 2-mm diameter, RGB metalens (designed under LPA) has an average focusing efficiency of 15\%~\citep{Li:22} (see Table~\ref{tab:1}). 

\begin{figure}[htbp]
    \centering
    \includegraphics[width=\linewidth]{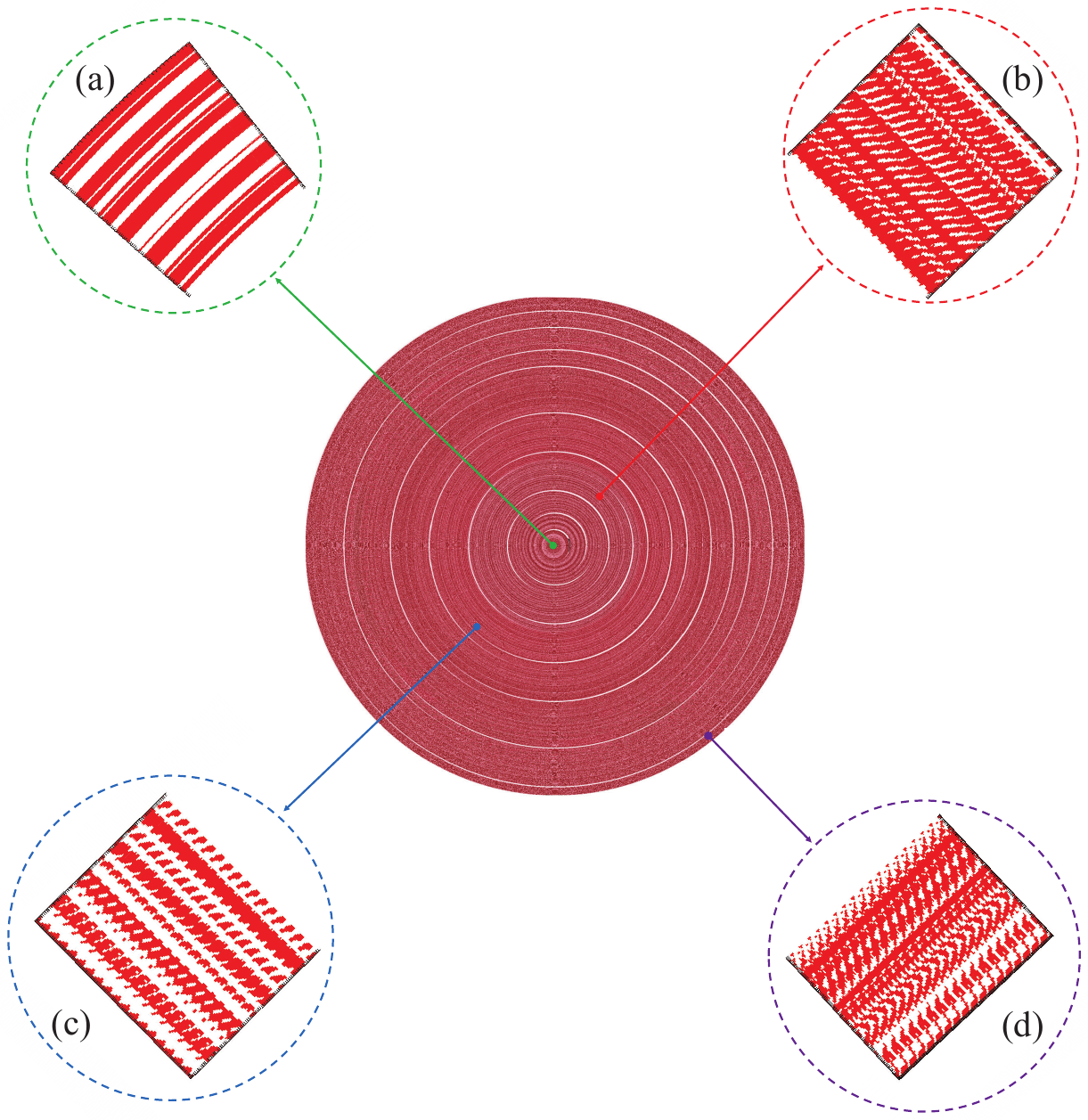}
    \caption{The schematic of the structure of the inverse-designed 2-mm diameter poly-achromatic metalens. Fig.~\ref{fig:7}(a) - (d) are the zoomed-in insets of the 1th (0-50$\lambda$), the 4th (150$\lambda$-200$\lambda$), the 8th (350$\lambda$-400$\lambda$) and the 16th (750$\lambda$-800$\lambda$) zones respectively.}
    \label{fig:7}
\end{figure}

\begin{figure}[htbp]
    \centering
    \includegraphics[width=\linewidth]{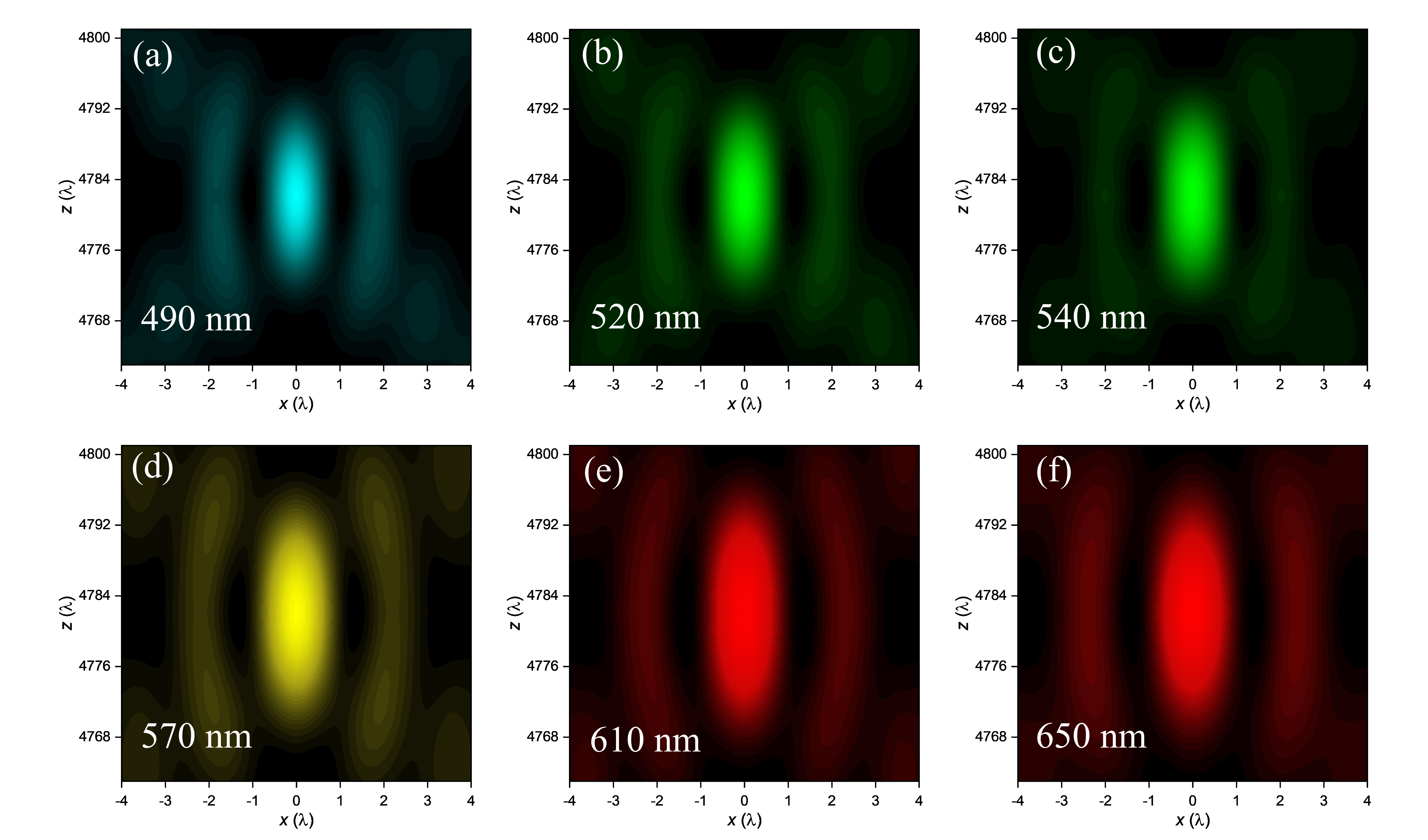}
    \caption{Simulated focal intensity distribution of the 2-mm-diameter and NA=0.3 poly-achromatic metalens in the $xz$-plane at (a) 490 nm, (b) 520 nm, (c) 540 nm, (d) 570 nm, (e) 610 nm and (f) 650 nm.}
    \label{fig:8}
\end{figure}

Secondly, we inverse-designed a 1.96mm diameter (3008$\lambda$) poly-achromatic (490nm, 520nm, 540nm, 570nm, 610nm and 650nm, corresponding to 0.75$\lambda$, 0.8$\lambda$, 0.83$\lambda$, 0.88$\lambda$, 0.94$\lambda$ and $\lambda$) metalens with a NA = 0.3 (corresponding to a focal length of 3.1mm or 4782$\lambda$). Again we use 650nm-thick titanium oxide on fused silica as the material platform.
The metalens is radially decomposed into 32 supra-$\lambda$ zones, each with a radial width of 47$\lambda$. Continuous axial symmetry is applied in the first zone while the azimuthal period for for the $\ell$-th zone ($\ell = 2, ..., 32$) is set to be $\tau^{(\ell)}_\phi = 2\pi / 1440\ell$. The spatial discretization step is set to be ${\lambda \over 36} = {\lambda \over 15 n_{index}}$. A conic filter with a radius of 0.056$\lambda$ is also applied. The optimized design is shown in Fig.~\ref{fig:7}. The far-field intensity distributions in the $xz$-plane at the six target wavelengths are shown in Fig.~\ref{fig:8}(a)-(f), which again shows a perfect achromaticity. The optimized metalens exhibits absolute focusing efficiencies of 12.8\%, 11.3\%, 10.5\%, 12.1\%, 12.9\% and 12.1\% at 490nm, 520nm, 540nm, 570nm, 610nm and 650nm respectively, while the average focusing efficiency is found to be 12\% overall. By comparison, the state-of-the-art, 2-mm diameter, poly-achromatic metalens (optimized for the same 6 wavelengths under LPA) has an average focusing efficiency of 8\%~\citep{Li:22} (see Table~\ref{tab:1}). 

Lastly, we report that the current generation of our software is able to finish a poly-achormatic freeform metalens design in about a week. Specifically, it took 156 hours on 8 Nvidia A100 GPUs to complete the design of 1.05-mm diameter RGB lens, and it took 214 hours on 8 Nvidia A100 GPUs to complete the design of 1.96-mm diameter multi-chrome lens. It is important to emphasize that a freeform topology optimization involves several thousands iterations, requiring a careful tuning and ramping up of binarization and conic filters to achieve a maximally performant design which is also manufacturable. To provide a better sense of the speed of our software, we also report that one iteration of the 1.96-mm design, for example, which extracts all the desired frequency responses at once in just one forward and one adjoint simulation, finishes in 12 minutes. Remarkably, this was achieved with only a preliminary GPU implementation using a limited number (8) of A100 GPUs confined to a single compute node, which resulted in a low GPU occupancy~\citep{Mei:23} and a moderate FDTD throughput of one billion voxels per sec per GPU, indicating that there remains substantial room for speedups (up to $30\times$ on a single GPU and more on multiple GPUs~\citep{Minkov:24}).   

Table~\ref{tab:1} summarizes state-of-the-art metalens designs, along with our results. Importantly, it should be noticed that millimeter or centimeter-scale ($\sim1000\lambda$), \textit{full-3D}, poly-achromatic designs are relatively rare (even under LPA), while a freeform design beyond LPA with such scales and characteristics is simply non-existent. Our results represent the first topology-optimized, $\gtrsim3000\lambda$-diameter, 3D freeform metalens designs without the drawback of LPA.

\section{Summary and Outlook}
\label{sec:s&o}
In this work, we demonstrated an adjoint-differentiable FDTD software for freeform topology optimization of millimeter and centimeter scale 3D metalenses. Our approach introduced the concept of ``zoned discrete axisymmetry'' (ZDA) enabling linear scaling with diameter coupled with freeform patterns, GPU acceleration, frequency-parallel broadband adjoint analysis, and adjoint-AD integration to achieve wide-aperture poly-achromatic metalens designs which significantly outperform the state-of-the-art. Even greater speedups are expected (via in-depth GPU occupancy maximization) in the next generation of our software in the near future. 

The broadband performance of a single-layer metasurface is inherently limited by its thickness, even with advancements in designing arbitrarily freeform nanostructures~\citep{Miller:23,Li:22(light),presutti:20}. However, our approach looks beyond ``simple'' metalenses---broadband or otherwise---by providing a critical tool which will empower the development of \textit{hyper-scale} computational metaoptics platforms. These platforms will incorporate \textit{large-area}, 3D volumetric nanophotonics~\citep{Lin:21(APL)} as well as leverage end-to-end \textit{co-}optimization of \textit{ultra-wide aperture} metaoptics hardware with advanced image-processing software~\citep{Lin:21(Nanophot),Lin:22(OE)}. In particular, the recent ``end-to-end'' co-design of metasurfaces for subsequent computational inference~\citep{Lin:21(Nanophot),Lin:22(OE),Arya:24,Fisher:22}, rather than for traditional image formation, has led to metasurface designs that are more and more different from the Fresnel-like patterns of previous metalenses, and such inference can even favor randomized surface patterns~\citep{antipa2017diffusercam,Li:23}---in such cases, LPA and similar approximations become increasingly unsuitable, and scalable full-wave simulations are essential. This transformative framework will, in turn, unlock unprecedented scales and complexities in the \textit{holistic} design of nanophotonics-enhanced computational imaging and computer vision systems, rivaling the scale and sophistication of celebrated Large Language Models. Furthermore, our scalable cylindrical-FDTD optimization software is not restricted to just metaoptics designs but will pave the way for a new generation of innovative architectures in both free-space and integrated (on-chip) large-scale photonics. Promising applications include novel discrete axisymmetric ring-resonator architectures for arbitrary nonlinear frequency-comb shaping~\citep{Lucas:23}, ultra-sensitive nonlinear multi-resonant gyroscopy~\citep{Sun:24} and dynamic obstacle-avoidance spatio-temporal beam-formation~\citep{Zhao:20}.  

\section{Materials and methods}
\subsection{Software development}
The software, including all key features discussed in the main text (the FDTD solver, the automatic differentiation environment and the GPU acceleration), have been developed using the Julia Language 1.10.4.

\subsection{Metalens design}
The inverse designs of the large-scale metalenses have been performed in Virginia Tech - Advanced Research Computing (VT-ARC). Using 8 Nvidia A100 GPUs, all three designs were finished in 150 - 220 hours. 

\begin{acknowledgement}
MS, AS, AK, DG, and ZL were supported in part by the US Department of the Navy (DON) under Contract No. N6833523C0545, the US Army Research Office (ARO) under Contract No. W911NF2410390 and the US Department of Energy under Contract No. DE-SC0024223.  SGJ was supported in part by ARO Award No. W911NF-23-2-0121 through the Institute for Soldier Nanotechnologies and by the Simons Foundation. QW and WTC were supported in part by the DON Contract N6833523C0545. We extend our gratitude to the TPOCs, Richard LaMarca and Chandraika Sugrim, for their invaluable support and guidance throughout this project under the DON Contract N6833523C0545.

% Please use ``The authors thank \ldots'' rather than ``The
% authors would like to thank \ldots''.

% The author thanks Mats Dahlgren for version one of \textsf{achemso},
% and Donald Arseneau for the code taken from \textsf{cite} to move
% citations after punctuation. Many users have provided feedback on the
% class, which is reflected in all of the different demonstrations
% shown in this document.

\end{acknowledgement}

\begin{suppinfo}

% A listing of the contents of each file supplied as Supporting Information
% should be included. For instructions on what should be included in the
% Supporting Information as well as how to prepare this material for
% publications, refer to the journal's Instructions for Authors.
We have included the test results of the simulation time of 3d axisymmetric metalenses with different radius in the Supporting Information. We have also attached the structural data (the $\varepsilon$ profiles of the topology-optimized discrete axisymmetric domains) for all of our metalens designs.

The following files are available free of charge.
\begin{itemize}
  \item permittivity - Centimeter-scale metalens at long-wave infrared.xlsx: The permittivity weight distribution of a centimeter-scale metalens at long-wave infrared
  \item permittivity - 1mm diameter RGB-achromatic metalens with NA = 0.8.xlsx: The permittivity weight distribution of a 1mm diameter RGB-achromatic metalens with NA = 0.8
  \item permittivity - 2mm diameter poly-achromatic metalens with NA = 0.3.xlsx: The permittivity weight distribution of a 2mm diameter poly-achromatic metalens with NA = 0.3
\end{itemize}

\end{suppinfo}

\bibliography{achemso-demo}

\end{document}